\documentclass[aps,amsmath,amssymb,prl,twocolumn,showpacs,superscriptaddress]{revtex4}
\usepackage{bm}
\usepackage{epsfig}
\usepackage[usenames]{color}

\begin{document}

%%%%%%%%%%%%%%%%%%% draft version display %%%%%%%%%%%%%%%%%%%%%%%%
%\begin{picture}(0,0)(0,0)
%\put(450,20){\makebox(0,0)[t]{\textcolor{red}{\bf version 9.5}}}
%\end{picture}
%%%%%%%%%%%%%%%%%%%%%%%%%%%%%%%%%%%%%%%%%%%%%%%%%%%%%%%%%%%%%%%%%%

\title{Frequency-Temperature Crossover in the  Conductivity of  Disordered Luttinger
Liquids}
\author{Bernd Rosenow}
\affiliation{Physics Department, Harvard University, Cambridge, MA
02138, USA} \affiliation{Institut f\"ur Theoretische Physik,
Universit\"at zu K\"oln, 50937 K\"oln, Germany}
\author{Andreas Glatz}
\affiliation{Materials Science Division, Argonne National
Laboratory, Argonne, IL 60439, USA} \affiliation{Institut f\"ur
Theoretische Physik, Universit\"at zu K\"oln, 50937 K\"oln,
Germany}
\author{Thomas Nattermann}
\affiliation{Institut f\"ur Theoretische Physik, Universit\"at zu
K\"oln, 50937 K\"oln, Germany}

\date{\today}

\begin{abstract}

The temperature ($T$) and  frequency ($\omega$)  dependent
conductivity of weakly disordered Luttinger liquids is calculated in
a systematic way both by perturbation theory and from a finite
temperature renormalization group (RG) treatment to leading order in
the disorder strength. Whereas perturbation theory results in
$\omega/T$ scaling of the conductivity such scaling is violated in
the RG traetment. We also determine the non-linear field dependence
of the conductivity, whose power law scaling is different from 
that of temperature and frequency dependence.

%Due to the strong effect of
%interactions in one spatial dimension, elementary charge excitations
%of a Luttinger liquid are plasmons. A random impurity potential
%leads to scattering of electrons and thus creates single particle
%excitations, whose interaction with plasmons gives rise to a
%pronounced energy dependence of backscattering. This interesting
%physics is observable in charge transport, whose {\it temperature,
%frequency, and electric field dependence} we discuss

\end{abstract}


\maketitle

Interacting one--dimensional electron systems display a large variety
of unusual and interesting phenomena, since not only interactions but
also external potentials (periodic or random) and thermal fluctuations
have pronounced effects on the behavior of these systems
\cite{Giamarchi03,gruener-rmp88}. The hallmark of
interaction effects in 1d systems is the power law single particle
density of states observable in the backscattering from a single
impurity  in a  Luttinger liquid (LL) \cite{KaneFisher}. While
the LL spectral function  has been observed experimentally \cite{spectral},
the rich behavior of collective scattering by random impurities has
eluded experimental observation so far.

In this letter, we focus on the effect of many weak (Gaussian)
impurities in a 1d disordered system. For noninteracting electrons,
this problem can be solved exactly \cite{Berezinskii73}. Interaction
effects are mostly treated perturbatively and described by a
dephasing length, which cuts off interference corrections. While
this regime of weak interactions has  been studied thoroughly
\cite{BeKi94,GoMiPo05}, less is known about the opposite regime of strong
interactions.  For attractive interactions, an unpinning transition
as a function of the interaction strength was found
\cite{fukuyama-lnp84,giamarchi+epl87,FuNa93}. In addition, the power
law exponent describing the energy dependence of impurity scattering
was predicted to flow as a function of energy
\cite{giamarchi+epl87}.  Finite temperature effects were partially
incorporated by truncating the renormalization group (RG) flow at
the de Broglie wave length of plasmon
excitations~\cite{giamarchi+epl87}.  However, for a complete study
of the thermal to quantum crossover, quantum and thermal
fluctuations have to be considered on an equal
footing~\cite{chakravarty+prl88}.

We calculate the frequency,  temperature, and electric field
dependence of the conductivity for spinless 1d fermions to leading
order in the disorder strength but for arbitrary short range
interactions. We go beyond previous approaches
\cite{fukuyama-lnp84,giamarchi+epl87,FuNa93,Giamarchi91} in several
respects. From a technical point of view, we i) present a systematic
approach to calculate the conductivity from a bosonic self energy in
the spirit of \cite{OsAf02}, both in perturbation theory and from a
finite temperature renormalization group
\cite{glatz+prl02,glatz+prb04}. This approach can be generalized to
higher orders in the impurity strength to obtain weak localization
corrections. We ii) explain that due to a
symmetry property of the self energy, an imaginary time RG
\cite{giamarchi+epl87,glatz+prl02} has a ballistic density
propagator although the retarded density propagator is diffusive.
From a physics point of view, we iii) present results for the
frequency-temperature crossover including the renormalization of the
interaction strength, and show that simple $\omega/T$-scaling is
violated in the RG solution.
 In addition, we iv) calculate the nonlinear field
dependence of the conductivity, which is characterized by a novel
power law exponent.  Our theoretical predictions  for weak disorder
are in reach of  present experimental technology, for instance in
carbon nanotubes \cite{TaSuWa00,CuZe04,TzChYi04} and polydiacetylen
\cite{AlLeCh04}, where the influence of strong disorder has already
been observed.  
Especially promising experimental systems are bent
2d electron systems in a strong magnetic field \cite{Gr+05} or
quantum Hall line junctions \cite{yang+04}, in which both
interaction  and disorder strength  can be  tuned.

{\it Model}.--- We consider particles moving in a random potential
$U(x) = \sum_i U_i \delta(x - x_i)$ due to point scatterers with
random positions $x_i$ and density $n_{\rm imp}$. The random potential
strength $U_i$ has moments $\overline{U_i}=0$ and $\overline{U_i U_j}
= U_{\rm imp}^2\delta_{ij}$. Throughout the paper,  we use units with
$\hbar=1$, $k_B=1$.  For noninteracting electrons with
Fermi velocity $v$, one finds a mean free path $1/\ell_0 = 2 n_{\rm
imp} U^2_{\rm imp}/ v^2$ by using standard perturbation theory. We
will use $\ell_0$ as an abbreviation for the disorder strength also in
the interacting case.  In the replica formalism, a system of spinless
interacting 1d electrons is described by the bosonic action
%
%******************************  action ************************************
\begin{eqnarray}
{\cal S}^{(n)}&  = &  \sum_{\alpha, \beta}
\int\limits_0^{L}dx\int\limits_0^{1/T}d\tau\Bigg\{ \frac{v}{2\pi
K} \Big[(\partial_x\varphi_{\alpha})^2+ {1 \over v^2}
(\partial_{\tau}\varphi_{\alpha})^2\Big]
\delta_{\alpha\beta} \nonumber\\
& & \hspace*{-.5cm}
-{1 \over \ell_0} { v^2 \Lambda^2 \over 8 \pi^2}
 \int\limits_0^{1/T}d\tau^{\prime} \cos{p}
\Big(\varphi_{\alpha}(x,\tau)-\varphi_{\beta}(x,\tau^{\prime})\Big)\Bigg\} \, .
\label{action.eq}
\end{eqnarray}
%***************************************************************************
%
Here, $\Lambda$ denotes the  cutoff in momentum space and the
interaction strength is parameterized by  $K$, with $K<1$ for
repulsive and $K>1$ for attractive interactions.
 The smooth part of the density is described by ${1 \over
\pi}\partial_x \varphi(x)$, giving rise to a forward scattering
term. Although it strongly influences static correlation
functions~\cite{glatz+prb04}, it has no effect on dynamic ones and
is hence neglected in this analysis.
 For CDWs, we have $p=1$, and $p=2$ for
spinless Luttinger liquids. In the following, we let $p=2$, and the
formulae for $p=1$ can be obtained by replacing $K \to K/4$.
  The optical conductivity can be calculated from
the retarded boson Green function as
%
%**************************  conductivity formula  **************************
\begin{equation}
\sigma(q,\omega) = e^2  \kappa  \ {- i \omega \over  q^2  - {\omega^2
\over v^2} +  \pi^2 \kappa  \Sigma^R(q,\omega)} \ .
\label{linearresponse.eq}
\end{equation}
%*********************************************************************
%
Here, the compressibility $\kappa = {K \over \pi v }$ is used as a
generalized density of states for interacting systems. All disorder
effects are contained in the retarded self energy $\Sigma^R$, which
is related to the mean free path via $1/\ell = - \pi K {\rm Im}[\Sigma^R/\omega]$.

{\it Perturbation theory}.--- We use the cosine-term in
Eq.~(\ref{action.eq}) as a vertex and keep in each term of its power
series two field operators as external operators and contract all
other operators with respect to the quadratic part of
Eq.~(\ref{action.eq}). This amounts to replacing  the cosine by  its
 best quadratic approximation. The  perturbative self energy is found to
be
\begin{figure}[t]
\includegraphics[width=0.9\linewidth]{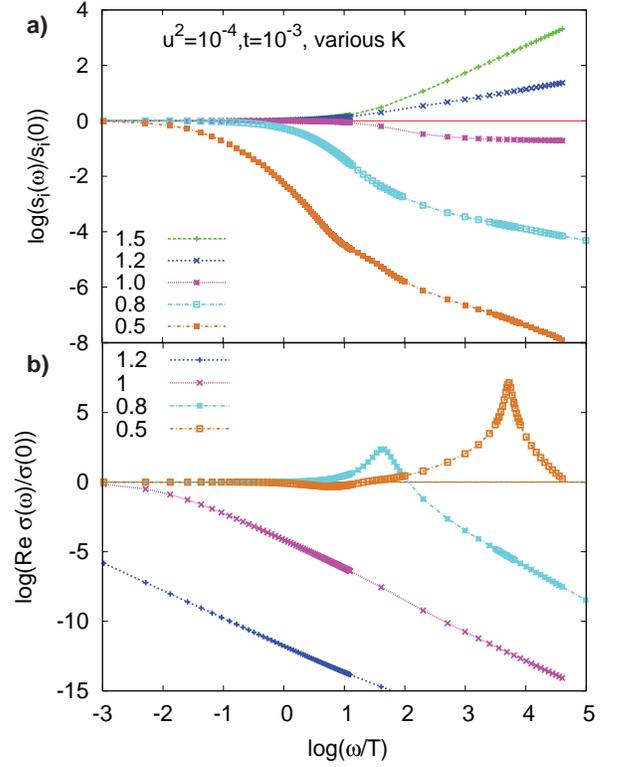}

\caption{a) Imaginary part $s_i={\rm Im}(\Sigma/\omega)$ of the self
energy calculated from finite temperature RG for several values of
$K_0$ (shown in the plots) at fixed $u_0^2=10^{-4}$ and $t_0=10^{-3}$.\\
 b) Double logarithmic plot of the real part of the conductivity
 calculate from Eqs.~(2), (10) for the same parameters,
rescaled to $1$ at $\omega=0$. }\label{wdep.fig}
\end{figure}
%
%**********************  perturbative self energy  *******************
\begin{equation}
\Sigma^{\rm per}_{\alpha \beta}(\tau)  =  -
\frac{v^2 \Lambda^2}{ \pi^2\ell_0  } \delta_{\alpha \beta}\Big[
e^{-{2} G_0(\tau)}
   - \delta(\tau)  \! \!  \int \limits_0^{1/T} \! \!  d\tilde{\tau}\,
e^{- 2 G_0(\tilde{\tau}) } \Big] \ .
\end{equation}
%*********************************************************************
%
For the bare local bosonic Matsubara Green function we use the expression
%
%*********************  bosonic Green function   ****************************
\begin{equation}
G_0(\tau) = K \ln \Big(1 + \Big|{\sin( \pi T \tau) \over \pi T
/\omega_c}\Big| \Big)  \label{greenfunction.eq}
\end{equation}
%*****************************************************************************
%
with $\omega_c=\Lambda v$. Proper regularization of the Green
function is especially important in the RG approach, where the
cutoff  flows to zero and the Green function needs to be evaluated
for energies larger than the cutoff. We calculate the retarded self
energy via the analytic continuation $\Sigma^R(t) = - 2\ \Theta(t)\
{\rm Im}\ \Sigma(\tau \to i t) $ with $\Theta(t)$ denoting the
Heavyside step function. After Fourier transforming, one obtains a
 scaling form  of the perturbative self energy  
%
%*************************** frequency retarded self energy  ******************
\begin{equation}
\pi^2 \kappa \Sigma^{R,{\rm per}}_{\alpha,\beta}(\omega) = -
\delta_{\alpha \beta} {} \frac{ \Lambda}{\ell_0} \Big({\omega\over
\omega_c} \Big)^{2 K-1}   F\Big({ \omega \over 2 \pi T}\Big) \ \ ,
\label{sigma_per.eq}
\end{equation}
%*******************************************************************************%
%
where the scaling function $F$ is given by
\begin{figure}[t]
\includegraphics[width=0.8\linewidth]{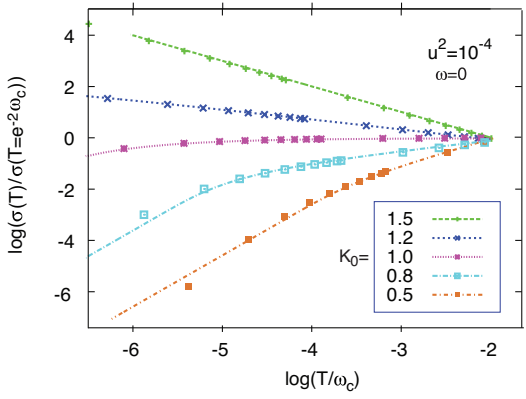}
\caption{Temperature dependence of the dc conductivity in log-log
representation for several values of $K_0$  and $u_0^2=10^{-4}$.
Lines: Results obtained by truncation of the zero temperature RG
equations (\ref{eq.dK/dl}-\ref{eq.du/dl}) at the thermal de Broglie
wave length (cf. Ref. [\onlinecite{giamarchi+epl87}]). Symbols:
Conductivity calculated from finite temperature RG and the
renormalized self energy (\ref{deltaSigmaR.eq}).}\label{tdep.fig}
\end{figure}
%
%************************* scaling function  ************************
\begin{eqnarray}
F(x) &  =&  - 4 x^{1-2K}K^2   \Gamma(- 2K)  \cdot
\label{scalingfunction.eq}\\
& & \hspace*{-1.5cm} \cdot \Big[ {| \Gamma(1 - K - i x)|^{- 2} \over
\cosh( \pi x) - i \cot( \pi K) \sinh(\pi x)} - \Gamma^{- 2}(1 -  K)
\Big]  \ .    \nonumber
\end{eqnarray}
%********************************************************************
%
In the derivation of Eq.~(\ref{scalingfunction.eq}), the infrared limit
$T/\omega_c \ll1$, $\omega/\omega_c \ll 1$ was taken and the 
regularization in Eq.~(\ref{greenfunction.eq}) was neglected.
The imaginary part of this scaling function is a good approximation
for all values of $K$, its real part  only  for $K \lesssim 0.5$;
 however, for $K
\approx 1$ and larger, the real part of $\Sigma$ can be neglected
compared to the $\omega^2$ term in Eq.~(\ref{linearresponse.eq}).
Interestingly, the part $ i \cot( \pi K) \sinh(\pi x)$ giving rise
to the imaginary part of the self energy vanishes identically for
all $x= i n$ with integer $n$, i.e. it does not appear in the
Matsubara self energy. The scaling function has the limiting forms
${\rm Im}[F(x)] \sim  i  $ for $x \to \infty$ and ${\rm Im} [F(x)] \sim  i x^{2-2K}$ for
$x\to 0$. Thus, interactions renormalize the noninteracting mean
free path $\ell_0$ with the $(2 - 2 K)$-th power of energy. The
perturbative expression Eq.~(\ref{scalingfunction.eq}) is valid for
${\rm max}[{\omega \over \omega_c}, {T \over \omega_c}] >  {1 \over
\ell \Lambda}$ \cite{bath} with the renormalized mean free path $\ell
=\Lambda^{-1}(l_0\Lambda/K^2)^{1/(3-2K)}$. Using the asymptotic
behavior of $F(x)$, we find for the real part of the conductivity
%
%***************  real conductivity  *******************
\begin{eqnarray}
{\rm Re} \; \sigma \! \!& \approx & \sigma_0\,\,\,\! \! \! \left\{
\begin{array}{ll}  \! \!  { ({2\pi  T \over \omega_c })^{2(1-K)}
\    {\Gamma(2K) \over  \Gamma^2( K) }\over
1 + ({\omega \ell_0 \over v})^2 ({2 \pi T \over \omega_c})^{4 - 4K} 
{\Gamma^2(2K) \over  \Gamma^4( K) }},
&   \! { \omega \over T} \ll1
\label{sigmaapprox.eq}\\[.5cm]
 {  ({\omega \over \omega_c })^{-2(2-K)} }
\
  {K \over \Gamma(2K)},
&   \!   { \omega \over   T} \gg 1
\end{array} \right.
\label{sigmalimiting.eq}
\end{eqnarray}
%**********************************************************
%
Here,
$\sigma_0 = e^2 \kappa \ell_0 v$. For $K \lesssim 0.5$, in the
regime ${\omega\over T} \gg 1$  a peak at the {\em temperature
dependent} frequency $\omega_{\wedge}\approx {v \over \ell} (\ell_0
\Lambda)^{1 - 2 K \over 6 - 4K} ({\omega_c \over 2 \pi T})^{1 - 2
K}$ appears (see Fig.~\ref{wdep.fig}b). For $K=1$, the Drude
conductivity $\sigma = \sigma_0/(1 + \omega^2 v^2/\ell_0^2) $ is
recovered. 
 For $K > 3/2$, perturbation theory is valid for all
frequencies, and the conductivity is imaginary with a $1/\omega$
divergence characteristic for a superconductor.
\begin{figure}[t]
\includegraphics[width=0.85\linewidth]{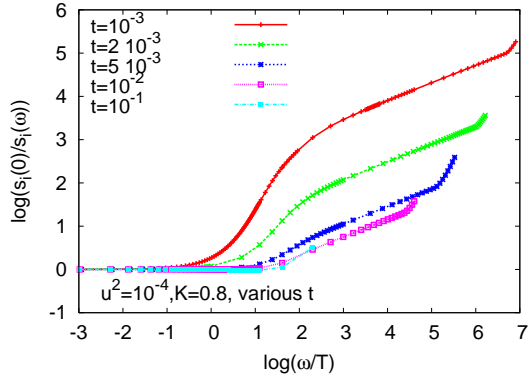}
\caption{Double logarithmic plot of the imaginary part $s_i = {\rm
Im}(\Sigma/\omega)$ of the self energy for $K=0.8$ and different
temperatures $t_0$.  As $K$ varies with frequency, curves for
different temperatures do not collapse and $\omega/T$-scaling is
violated.}\label{scaling.fig}
\end{figure}

{\it Renormalization}.--- Since the derivation of the flow
equations is well documented in the literature we will here quote
only the result \cite{glatz+prl02,glatz+prb04}. At finite
temperature, the flow equations are given by
%
%****************************  flow equations  ********************************
\begin{subequations}
\begin{eqnarray}
\frac{dK}{dl} & = & - 8 u^2KB\left(K,\frac{K}{2t}\right) \coth
\frac{K}{2t}
\label{eq.dK/dl},\\
\frac{du^2}{dl} & = & \Big[3  -
2 K \coth{\frac{K}{2t}}\Big]u^2
\label{eq.du/dl}, \\
B(K,y) & = & \int\limits_0^y \frac{\tau^2\,d\tau}{\left[1 + \left({2
y \over \pi} \sin{\pi \tau \over 2 y}\right)^2
\right]^{K}}\frac{\cosh(y-\tau)}{\cosh{y}}  . \label{function.eq}
\end{eqnarray}
\end{subequations}
%***************************************************************************
%
 Here, the
dimensionless temperature $t$ obeys the flow equation $\frac{dt}{dl}
=t$ with initial value $t_0 \equiv t(l=0) = T K/ \omega_c$, and the
dimensionless disorder strength $u^2$ has the initial value $u_0^2 =
{1 \over \ell_0 \Lambda}{K^2 \over 2 \pi}$. Calculating the self
energy by integrating over a momentum shell $\Lambda \exp(- \delta
l) < |k| < \Lambda$, we obtain
%
%***************************** delta Sigma  ******************************************
\begin{equation}
\delta \Sigma_{\alpha \beta} (\tau) = 2 K {\cosh[\omega_c
(\tau - {1 \over 2T})] \over \sinh({\omega_c \over 2 T} )} \,
\delta l \, \Sigma^{\rm per}_{\alpha \beta}(\tau)
\end{equation}
%*************************************************************************************
%
for $0 < \tau < 1/T$.  Due to the symmetry property $\delta
\Sigma_{\alpha \beta}(\tau) = \delta \Sigma_{\alpha \beta}(1/T -
\tau)$ in this interval, the low frequency expansion of the Fourier transform
$\delta \Sigma_{\alpha \beta}(\omega_n)$ starts with a term $O(\omega_n^2)$,
which renormalizes the corresponding term in the action and gives
rise to the flow of $K$ in the RG. The analytical continuation $i
\omega_n \to \omega + i \eta$ would not give rise to a diffusive
term in the boson propagator. Instead, one has to perform the
analytic continuation in the time domain.
 Although the diffusive part of $\Sigma^R$
does not feed back in the RG, it determines the conductivity
 in an essential way.  The incremental change of the
retarded self energy is
%
%***************************  delta Sigma retarded *****************************
\begin{widetext}
\begin{equation}
\delta \Sigma^R(\omega,T,l) =  {\Lambda^3 v^2 u^2e^{-3l}  2 \delta l
\over \pi K
 \sinh(v  \Lambda/2 T)}  \int \limits_0^\infty  \! \!   d\tilde{t}  {e^{i
 \omega \tilde{t}} -1 \over \Big[1 + {\sinh^2(\pi T \tilde{t}) \over (\pi
 T/v  \Lambda)^2}\Big]^{K}}\ \
 {\rm  Im}\Big\{ \exp\Big({2 K \over  i}  \arctan\Big[{\sinh(\pi T \tilde{t})
 \over \pi T/v  \Lambda} \Big] \Big) \cosh\Big(- {v  \Lambda \over 2 T} + i \tilde{t} v
 \Lambda\Big) \Big\}
\label{deltaSigmaR.eq}
\end{equation}
\end{widetext}
%*******************************************************************************%
%
The variables  $K$, $v=K/(\pi \kappa)$ and $\Lambda$ are functions
of the RG scale $l$ as determined from the flow equations
Eqs.~(\ref{eq.dK/dl}). Note that there is no renormalization of
$\kappa$.
%, (\ref{eq.du/dl}), whereas $u_{\rm eff}$ is the renormalized but
%unrescaled disorder strength.
The retarded self energy to be used
in the conductivity Eq.~(\ref{linearresponse.eq}) is obtained by
integrating over the flow parameter $l$ from zero to infinity. The
frequency-temperature crossover of self energy and conductivity is
displayed in Fig.~\ref{wdep.fig}. One clearly sees that the power
law exponent of the frequency dependence changes with energy.
Besides the scale dependence of exponents, the conductivity is
characterized by the limiting form described in
Eq.~(\ref{sigmalimiting.eq}) and the following discussion. The
temperature dependence of the dc conductivity is compared to the
approximate result \cite{giamarchi+epl87} in Fig.~\ref{tdep.fig}. We
observe qualitative agreement between the two approaches. Due to the
flow of the Luttinger parameter $K$, $\omega/T$-scaling of the self
energy is violated (see Fig.~\ref{scaling.fig}).
\begin{figure}[h]
%\vspace*{-1cm}
\includegraphics[width=0.85\linewidth]{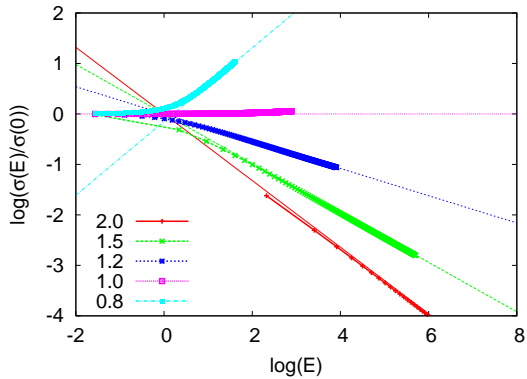}
\caption{Nonlinear field dependence of the inverse mean free path
${1\over \ell} \sim E^{2 K -2\over 2 K -1}$  for different values of
$K$. The straight lines represent the power
law.}\label{nonlinear.fig}
\end{figure}

{\it Nonlinear field dependence.--} In the regime of linear dc transport, the average excitation energy due to the external electric field   is $e_0 E \ell(T)$.  In
the nonlinear regime, the electric field  dependence $\ell(E,T)$ of the mean free path
must be taken into account. The {\em frequency} and
temperature dependence
 of $\ell$ can be directly calculated from the imaginary part of
the self energy via ${1 \over v \ell} = - \pi^2 \kappa {\rm Im}(\Sigma^R(\omega,T)/\omega)$. In order to obtain the {\em electric field}
and temperature dependence of $\ell$,
we recall that for scattering
from a single impurity in an LL, the nonlinear conductance
$G_{\rm nl}(V,T)$ and the ac conductance $G_{\rm ac}(\omega,T)$ are
related by $G_{\rm nl}(V,T) = G_{\rm ac}(e_0 V,T)$ to leading order in the impurity strength. By analogy, the nonlinear field dependence of the
mean free path can be obtained by replacing the frequency  in
the self energy by the average electron energy $e_0 E \ell(E,T)$.
In this way,   $\ell(E,T) $ can be found as the solution of the
equation
%
%*****************  interacting mean free path  *******************
\begin{equation}
{1 \over \ell(E,T)}  = - \pi K {\rm Im}\left[ {\Sigma(e_0 E \ell(E,T),T)
\over e_0 E \ell(E,T)} \right] \ .
\label{selfconsistency.eq}
\end{equation}
%*******************************************************************
%
Using the perturbative expression Eq.~(\ref{sigma_per.eq}), this
condition simplifies to ${\ell(E) \over \ell_0} = \Big({e_0 E
\ell(E)\over \hbar \omega_c}\Big)^{2 - 2 K}$ for temperatures
$T \ll e_0 E \ell(E)$,  and one obtains for the
nonlinear dc conductivity
%
%********************* nonlinear dc conductivity  ****************
\begin{equation}
\sigma_{\rm per}(E) \approx  \sigma_0 \
\Big({e_0 E \ell_0
\over \hbar \omega_c}\Big)^{2 - 2K \over 2 K -1} \ \ .
\label{sigma_noninear.eq}
\end{equation}
%******************************************************************
%
This zero temperature approximation is valid only for $K >  {1 \over
2}$, as for lower values of $K$ a physical solution to
Eq.~(\ref{selfconsistency.eq}) can be found only at finite
temperature. In Fig.~\ref{nonlinear.fig}, $\sigma_{\rm nl}(E,T)$ as
calculated from the perturbative self energy
Eq.~(\ref{sigma_per.eq}) is shown.

{\it Conclusions}.--- We have discussed frequency, temperature, and
field dependence of  charge transport in a disordered  LL in a
bosonized theory. The conductivity is obtained from the bosonic self
energy, which is integrated from the flow equations of a finite
temperature RG. In contrast to single impurity physics in a LL, the
power law exponent describing impurity scattering is scale dependent
in the disordered case. The  mean free path in nonlinear dc transport  
 is selfconsistently 
calculated  by replacing
the photon energy in the ac self energy by the average electron
energy acquired between  scattering events.

{\it Acknowledgments:} We thank T.~Giamarchi and B.~Halperin for
helpful discussions. B.R. acknowledges support from a Heisenberg and
A.G. from a research stipend of  DFG.


\begin{thebibliography}{10}
\bibitem{bath} For ${\omega  \over \omega_c} \ll {1 \over \ell \Lambda}$ and weak interactions $K \approx 1$,
 weak localization corrections are  important even  for ${T \over \omega_c}$
 parametrically larger than ${1 \over \ell \Lambda}$, 
 see   \cite{GoMiPo05}.
\end{thebibliography}
\end{document}